# Pourquoi j'ai tué mon cuivre – mise en lumière de la FTTO dans l'ESR

**Gabriel Moreau**
Laboratoire LEGI – UMR5519
1209-1211 rue de la piscine
38240 Saint-Martin d'Hères

**Bernard Maire-Amiot**
Institut Néel – UPR 2940
25 rue des Martyrs
38000 Grenoble

**David Gras**
CNRS Délégation Alpes – DR11
25 rue des Martyrs
38000 Grenoble

**Hervé Colasuonno**
Laboratoire G2ELAB – UMR 5269
21 avenue des martyrs
38000 Grenoble

**Julien Bamberger**
Laboratoire G2ELAB – UMR 5269
21 avenue des martyrs
38000 Grenoble

**Aurélien Minet**
École Pratique des Hautes Études
4-14 rue Ferrus
75014 Paris

**Alain Péan**
Laboratoire C2N – UMR 9001
10 boulevard Thomas Gobert
91120 Palaiseau

**Marie-Goretti Dejean**
CIRM – UMS 822
163 avenue de Luminy
13009 Marseille

## Résumé

*La **FTTO** [1] signifie **F**iber **T**o **T**he **O**ffice, au regard de la FTTH [2] (Fiber To The Home) déployée en France chez les particuliers. Le principe de la FTTO est de câbler un bâtiment tout en **fibre optique**, d'éliminer au maximum tout câble en cuivre et de poser des micro-commutateurs [3] (μsw) optique / cuivre dans chaque bureau, au plus* près des machines. La connexion des utilisateurs s'effectue toujours avec leur jarretière cuivre RJ45 [4] classique. Au travers de questions / réponses, nous mettrons en avant les raisons pour lesquelles **la FFTO est une technologie maîtrisée et d'avenir**.

*Durant ces six dernières années, plusieurs projets de bâtiments dans le périmètre de l'ESR ont choisi cette technologie, sont sortis ou vont sortir de terre. Selon le projet, différentes topologies et technologies sont possibles. Quels sont les retours d'expériences après ces années ? Le résultat est-il conforme aux attentes ? Comment vit la solution au jour le jour ? Quelle sécurité, comment configurer et maintenir un large ensemble de commutateurs, quelle haute disponibilité est possible ? Comment s'intègrent le Wi-Fi, la téléphonie sur IP ainsi que tous les équipements PoE [5] ? La FTTO participe-t-elle à limiter la consommation ?*

*Comment concrètement monter un appel d'offre FTTO pour un projet, quels sont les éléments essentiels à intégrer et quelles sont les erreurs à éviter à tout prix ? Et demain, quelle espérance de vie pour son infrastructure et quels débits envisageables ?*

*Le groupe FTTO [6] de RESINFO [7] travaille à donner **des réponses claires** à l'ensemble de ces questions ainsi qu'à faire **partager son expérience** auprès de la communauté.*

## Mots-clefs

*FTTO, Fibre optique, Topologie, Cheminement, Monomode, Réseau, Bâtiment, Câblage, Retours d'expérience, Gestion des configurations, Gestion de la puissance, Redondance des chemins, Boucle optique, Supervision, Cartographie, Boîtier de raccordement*

## 1 Introduction : pourquoi je vais devoir tuer mon cuivre

Le câblage réseau d'un bâtiment était jusqu'à récemment basé sur du cuivre. Avec l'augmentation des débits, ce type de câble est devenu de plus en plus gros, lourd, coûteux (prix de la matière première) et surtout plus technique (catégorie Cat-6a [8], Cat-7, voire Cat-8). Par ailleurs, plus la liaison cuivre est longue, plus l'affaiblissement du signal est important. Si cette atténuation existe également pour la fibre optique, elle est bien moindre et négligeable à l'échelle d'un bâtiment pour de la fibre monomode. Par conséquent, en cuivre, il faut poser un élément actif au moins tous les 100 m, ce qui augmente les frais occasionnés (locaux techniques…) et complexifie la gestion du réseau. De plus, le câble en cuivre étant sensible aux perturbations électromagnétiques, les installations réseaux doivent prendre en compte la proximité des lignes de courant fort, des cages d'ascenseur, des équipements (par exemple : expérimentations scientifiques) engendrant des courants ou des champs magnétiques forts.

L'arrivée de la technologie PoE conduit à l'augmentation de la consommation des commutateurs et en conséquence à une potentielle climatisation des locaux techniques. Les dernières évolutions de la norme PoE++ améliorent l'efficience énergétique, et les alimentations de type 4 injectent un maximum de 90 W (950 mA) en sortie de commutateur. Cependant, selon la longueur et la qualité du câble, la puissance à la prise n'est plus que de 70 W. Il faut donc essayer de limiter la longueur des câbles en cuivre afin d'éviter les pertes et leur échauffement. C'est pourquoi il faut également prêter attention au PoE++ au cœur des torons cheminant sur les chemins de câbles. À noter qu'avec EEPoE, technologie prenant en compte le PoE au niveau IEEE 802.3az [9], on peut espérer sauver 1 ou 2 W par port.

De son côté, la fibre optique, autrefois réservée aux longues distances et/ou aux forts débits en datacentre, est devenue de moins en moins chère, en particulier du fait du déploiement de la FTTH (Fiber To The Home) chez les particuliers qui en consomme des millions de km par an. Au point même que la fibre monomode, aux caractéristiques nettement plus performantes et adaptées aux longues distances, est devenue moins chère au km que la fibre multimode, qui était jusqu'ici systématiquement utilisée sur les courtes distances. Les prix des modules SFP [10] monomode, autrefois nettement plus chers que leurs équivalents multimodes, ont beaucoup baissé. Il existe actuellement des SFP LX duplex à moins de 20 € HT (à condition de ne pas prendre des *transceivers* [11] propriétaires). Considérée historiquement comme plus fragile que le cuivre, la fibre optique est devenue nettement plus résistante. Elle peut supporter un rayon de courbure nettement plus petit (spécialement la G657a [12]) et une traction beaucoup plus forte (les câbles sont renforcés au Kevlar). Avec son faible diamètre, un câble peut donc se déployer facilement là où il y a peu d'espace, traverser un mur porteur, se faire discret dans un monument historique ou dans le faux plafond, *etc*. La fibre optique est insensible aux perturbations électromagnétiques et peut donc être installée près d'équipements engendrant de telles perturbations (courant fort 20 kV, 400 V, ateliers, laboratoires…).

**Pourquoi ne pas profiter de « cette manne fibre » au plus près de nos postes de travail ?**

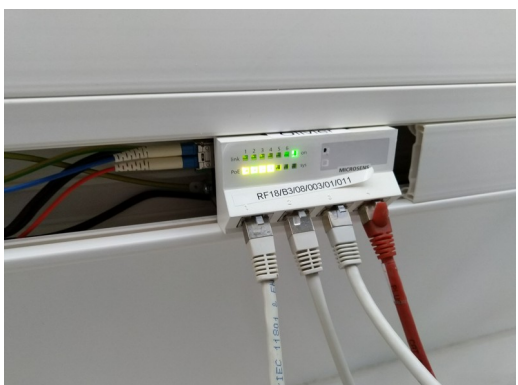

*Figure 1: Micro-commutateur en goulotte au C2N*

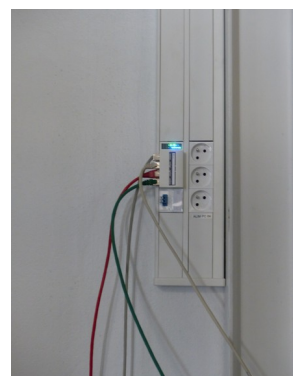

*Figure 2: Micro-commutateur en goulotte au LEGI en position verticale*

Actuellement, la majorité des équipements informatiques n'est pas dotée de port fibre et conserve des ports RJ45 cuivre classiques. De plus, les périphériques tels que les téléphones ToIP, les bornes Wi-Fi, les équipements de GTB [14] ou les caméras de surveillance nécessitent souvent une alimentation PoE. La solution retenue consiste donc à placer à l'extrémité des fibres optiques, des commutateurs ou micro-commutateurs, qui peuvent être en goulotte (solution préconisée par les deux leaders de la FTTO – figure 1 et 2). Ceux-ci exposent en façade et parfois en interne des ports RJ45 pouvant fournir du PoE.

En résumé, **la fibre arrive directement dans le micro-commutateur situé dans le bureau**, c'est-à-dire « **F**iber **T**o **T**he **O**ffice » : **FTTO**. Ainsi, tout équipement en cuivre, PoE ou non, peut être branché « classiquement » au micro-commutateur le plus proche (figure 3). La fibre permet aussi d'amener du très haut débit 10 Gb/s (SFP+) ou 25 Gb/s (SFP28) directement sur les machines en faisant un raccordement direct (figure 4).

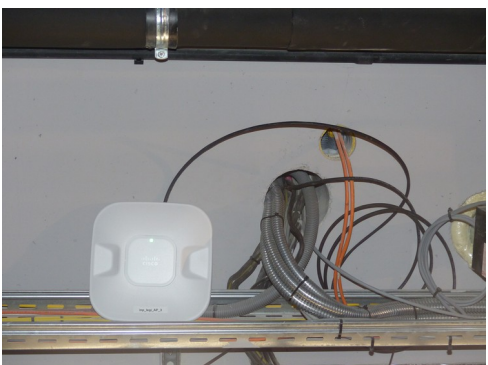

*Figure 3: Borne Wi-Fi connectée directement sur le port interne du micro-commutateur le plus proche*

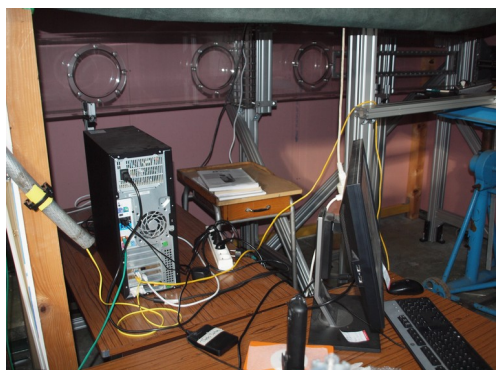

*Figure 4: Station d'acquisition directement connectée en fibre optique à 10Gb/s [13]*

La technologie FTTO est apparue en Allemagne au début des années 1980. En 2013, le LEGI (Laboratoire des Écoulements Géophysiques et Industriels) situé sur le Campus Universitaire de Grenoble a été le premier bâtiment universitaire de France câblé en FTTO. Voici un panel représentatif des établissements et des laboratoires équipés en FTTO (enquête RESINFO FTTO 2019) :

- Laboratoire LEGI, Grenoble, 2013 : 94 micro-commutateurs à ce jour [15] ;
- Grenoble Énergie – Enseignement Recherche (GreEn-ER / PPP [16]), Grenoble, 2015 : 800 micro-commutateurs [17] ;
- Institut du Génome Humain, Montpellier, 2016 : 80 mini-commutateurs[1] [18] ;
- ANR, Paris, 2016 : 300 micro-commutateurs ;
- Université Marne La Vallée, 2016 et suivantes : 1600 micro-commutateurs [19] ;
- École Centrale / Supélec, Saclay, 2017 : 2200 micro-commutateurs [20] ;

---

1. Il s'agit de mini-commutateurs ayant 8 ports RJ45. Ils sont fixés aux abords de la goulotte électrique et non sur celle-ci. Le port *uplink* SFP partage le débit de 1 Gb/s entre 8 machines maximum et non 4 dans le cas de micro-commutateur (téléphonie comprise). Cette solution est donc moins intégrée, légèrement moins performante, mais elle est financièrement plus abordable. Une plus grande variété de type de commutateurs peut être envisagée dans son infrastructure. À noter qu'un mix mini-commutateur / micro-commutateur est tout à fait possible dans une installation.

- Direction délégation régionale CNRS DR11 Grenoble, 2017 : 15 micro-commutateurs [21] ;
- Institut Néel, Grenoble, 2017, 2018 : 180 micro-commutateurs [22] ;
- Centre de Nanosciences et Nanotechnologies (C2N / Maîtrise d'ouvrage CNRS), Saclay, 2018 : 800 micro-commutateurs ;
- Centre International de Rencontres Mathématiques (CIRM), Marseille, octobre 2019.

**Remarques :**

- le système FTTO déployé dans les bâtiments ci-dessus n'a aucun rapport avec certaines solutions commerciales FTTO qui « amènent la fibre à une PME », ni avec la FTTH, la FTTB ou la FTTE [23] ;
- les dates indiquées renseignent sur l'année de prise en main des locaux. Les dossiers ont donc été préparés *a minima* 2 ans plus tôt, voire bien plus. Il y a donc une certaine expérience au niveau de ces déploiements ;
- d'autres sites de l'ESR sont actuellement en cours de réalisations (par exemple : projet 4R4 à Toulouse) ou sont encore à la phrase d'écriture et/ou de vérification du cahier des charges.

Cet article vise à synthétiser et vous présenter les diverses expériences vécues dans nos laboratoires afin de vous accompagner dans votre prochain projet de tuer votre cuivre. **Il est donc temps de passer de l'âge du cuivre à celui des lumières.**

## 2 Aide pour bien démarrer

En juin 2018, dans le cadre de RESINFO (Réseau de métier des ASR dans l'enseignement et la recherche), les Journées Systèmes (JoSy [24]) dont l'objet était la FTTO ont eu lieu à Grenoble. En parallèle un groupe de travail national s'est constitué dans le but :

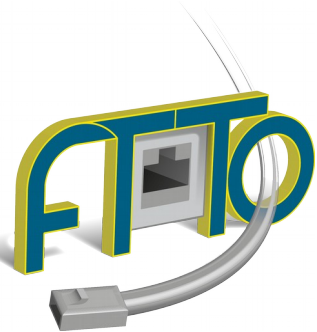

- d'échanger sur nos expériences sur ce sujet ;
- d'aider les laboratoires qui souhaiteraient se lancer dans cette aventure.

Nous devons apprendre de nos échecs et de nos réussites et en faire profiter notre communauté. De plus, rien ne vaut la visite de plusieurs sites équipés avec ce type de matériel. Il faut échanger avec l'ensemble des acteurs, que ce soit des personnels de la BAP E [25] (Informatique, Statistiques et Calcul scientifique) ou des personnels de la BAP G (Patrimoine immobilier, Logistique, Restauration et Prévention).

Il faut avoir l'esprit ouvert avant de démarrer un tel projet, ne pas avoir peur du changement, car avec la FTTO nous devons tout repenser : l'architecture, l'encombrement, la technologie, la connectique…

N'hésitez pas à nous contacter et à venir visiter nos sites en production, nous avons franchi le pas, alors vous pouvez le faire ! Si vous vous lancez, « *non, vous n'êtes pas* *tout seul »*, il existe une liste **ftto@listes.resinfo.org** qui est active et réactive. De même, la rubrique du groupe FTTO du site web de RESINFO s'étoffe avec le temps d'articles consacrés à la FTTO.

## 3 Architecture de l'infrastructure optique du réseau

Comme souvent dans les projets technologiques, la solution unique n'existe pas. Plusieurs solutions sont possibles selon le contexte local. Nous allons décrire ici l'architecture la plus déployée actuellement dans l'ESR.

La première chose qui marque lors d'une visite de site est la quasi-absence de chemin de câble et de câbles dans un cœur de réseau FTTO, comparativement à une solution traditionnelle en cuivre (figure 5 et 6). Nous allons comprendre pourquoi.

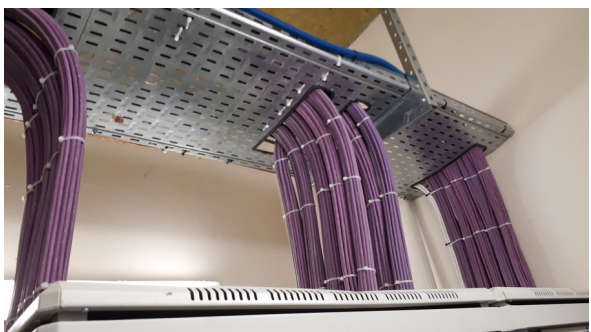

*Figure 5: Un des sous-répartiteurs en cuivre à l'Institut Néel – 450 terminaisons RJ45*

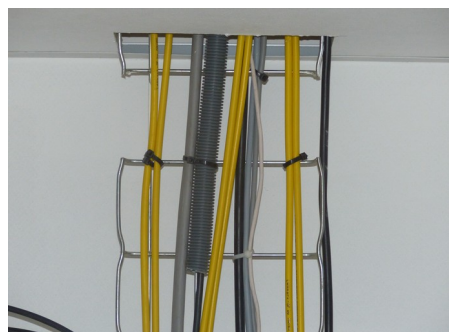

*Figure 6: Répartiteur central au LEGI – 6 câbles de 72 fibres (3 boucles) – 432 ports LC, soit 864 terminaisons RJ45 potentielles en duplex (1728 en simplex)*

N'étant pas limité par la distance, nous allons relier en point à point directement chaque bureau avec le cœur de réseau, sans chercher à prendre le chemin le plus court. Le prix du mètre linéaire de la fibre est très faible sur le coût global d'un projet. On déroule, dans les zones de cheminement, en faux plafond, un câble optique comportant N tubes de M brins. Pour la suite de l'article, nous allons prendre le cas classique d'un câble ayant 12 tubes de 12 fibres chacun, soit 144 fibres en tout. L'autre extrémité de ce câble arrive elle aussi dans le cœur du réseau en empruntant si possible un autre chemin au sein du bâtiment. On a ainsi créé **une boucle en fibre optique qui part et revient en salle serveur** (figure 6, 7 et 8). Pour le moment, peu d'intérêt ! Chacune des fibres est distribuée sur un panneau de brassage (figure 12 et 13). Nous avons donc au final 288 ports LC [26] (ou SC) sur le tableau central : les départs et les arrivées de la boucle. Une première recette optique est alors faite. Si le bâtiment est grand et/ou comporte plusieurs étages, plusieurs boucles peuvent être déployées, en partant toujours du même cœur de réseau.

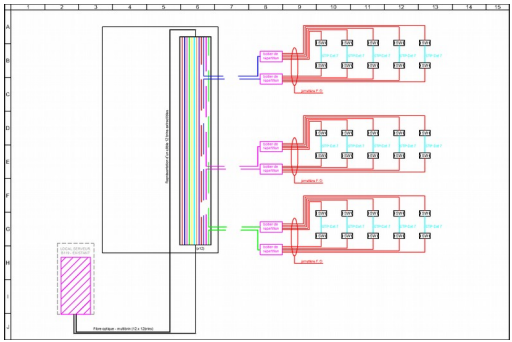
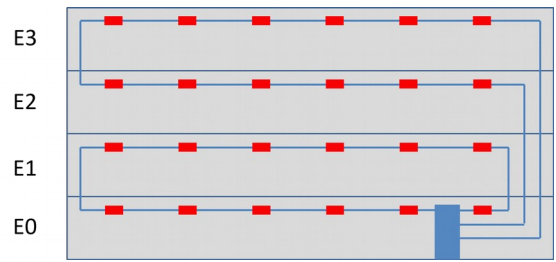

*Figure 8: Schéma de principe des boucles optiques sur un bâtiment en rénovation à l'Institut Néel*

*Figure 7: Schéma de principe du câblage à la DR11 (CNRS Grenoble Alpes)*

Afin de pouvoir **relier un bureau**, il va falloir **dériver une fibre** (figure 9). Cela se réalise en coupant intelligemment un tube. En pratique, à intervalles plus ou moins réguliers, selon la disposition des bureaux et des salles d'expérimentations, un boîtier d'éclatement est positionné sur le câble. Au niveau de celui-ci, en général 1 à 2 m avant ou après, on coupe un tube du câble, par exemple le tube bleu (chaque tube et chaque fibre a une couleur permettant de l'identifier). Il n'est en effet pas possible actuellement de couper juste une fibre.

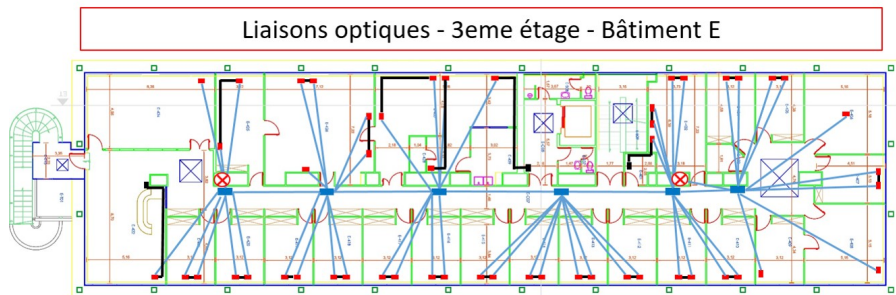

*Figure 9: Schéma de câblage détaillé sur un étage à l'Institut Néel*

Les 12 fibres optiques du tube en sont alors extraites, puis rallongées par des pigtails ayant pour extrémités des prises LC (figure 11). Par exemple, s'il a été choisi de couper le tube bleu après le boîtier, les 12 prises LC sont reliés en salle serveur aux 12 prises LC du tube bleu de l'origine de la boucle. Il est donc possible dans un autre boîtier positionné plus loin de couper le tube bleu avant celui-ci et donc d'utiliser les 12 fibres sur l'autre partie de la boucle optique (figure 10). Entre les deux boîtiers, le tube bleu ne sert plus à rien. Dans une structure en boucle, chaque tube, chaque fibre peut servir deux fois, dans un sens et dans l'autre, sur deux boîtiers différents (il est possible de le faire sur un même boîtier, solution partiellement déployé au LEGI, mais nous le déconseillons). Notre boucle est donc équivalente à un câble de 24 tubes au final ! Une seconde recette optique permettant de certifier la dérivation et chaque fusion de pigtails est réalisée entre toutes les prises LC d'un boîtier de raccordement et le panneau de brassage centralisé.

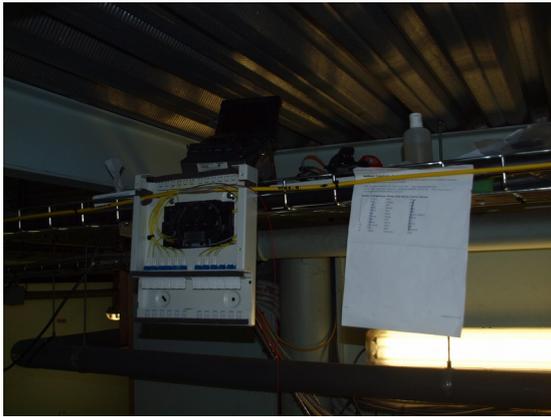 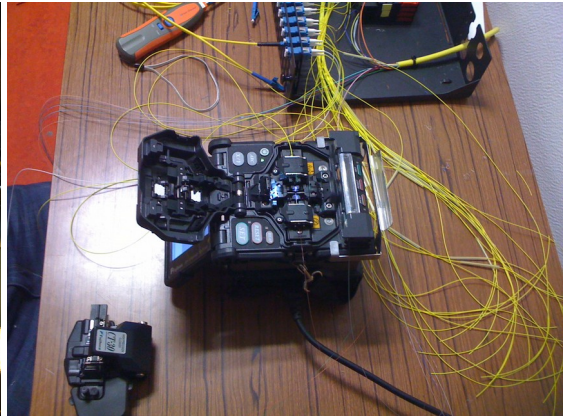

*Figure 10: Boîtier de raccordement posé à quelques centimètres d'un câble 400V, avec 3 tubes de 6 fibres coupés à 2 m à droite du boîtier, puis extraits du câble*

*Figure 11: Fusionneuse : fusion des pigtails LC sur les fibres d'un tube*

Afin de câbler un bureau, il suffit alors de poser une jarretière optique LC-LC de la longueur standardisée la plus adaptée (10, 15, 20, 25 m…) entre le boîtier de raccordement et l'emplacement final du micro-commutateur. Environ 1 m « de mou » est laissé dans la goulotte et le reste est enroulé près du boîtier, fixé sur la tranche du chemin de câble. Habituellement, chaque micro-commutateur dispose de 4 ports externes directement utilisables par les utilisateurs et d'au moins 2 ports internes, dont un au format SFP (d'autres variantes sont disponibles sur les catalogues constructeurs). Le commutateur est donc directement branché via un transceiver sur la fibre, elle-même directement reliée au cœur de réseau. Le micro-commutateur ayant 4 ports en façade avant, un seul port SFP en cœur de réseau alimente 4 ports utilisateurs. Il y a donc besoin dans ce cœur de réseau de quatre fois moins de ports qu'une solution centralisée tout en cuivre, la commutation étant partiellement décentralisée (figure 5, 12 et 13).

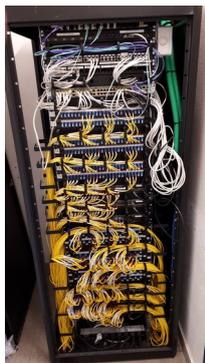 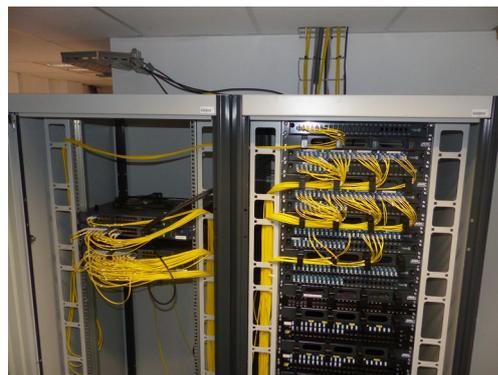

*Figure 12: Brassage par bâtiment en fibre optique à l'Institut Néel. Suppression de tous les sous-répartiteurs d'étage.*

*Figure 13: Brassage centralisé au LEGI. Point unique pour 432 ports fibres répartis sur 4 bâtiments.*

**Nous avons ainsi créé le réseau en étoile parfait en partant d'une boucle fermée sur elle-même !**

Une fois cette architecture bien comprise, plusieurs variantes sont possibles :

- déployer des câbles fibres pré-connectorisés entre le cœur et un boîtier de raccordement, donc sans boucle. C'est la solution choisie à l'Université Libre de Bruxelles en interne dans ses bâtiments. Elle peut être aussi intéressante pour relier une annexe indépendante n'ayant que quelques bureaux ;

- relier deux micro-commutateurs voisins en RJ45 via le second port interne avec un câble cuivre de faible longueur. On crée ainsi une boucle Ethernet devant être gérée par Spanning-Tree [27]. Si les deux commutateurs ont été branchés sur des boîtiers de raccordement différents, l'un allant vers l'origine de la boucle et l'autre vers son extrémité, alors l'architecture sera redondante à la rupture d'un chemin optique (panne de transceiver, dégradation du câble optique…) pour le prix d'un câble RJ45 de quelques mètres ;

- avoir deux cœurs de réseau et faire partir les boucles d'un cœur pour aller vers l'autre cœur. Si vous reliez vos micro-commutateurs entre-eux intelligemment, votre architecture est alors aussi tolérante à la panne d'un des deux cœurs de réseau ;

- avoir des prises optiques LC dans la goulotte au format 45x45 en plus d'un commutateur. Il est ainsi possible de relier certaines stations de travail directement en optique, au 10 Gb/s ou au 25 Gb/s à un prix abordable !

- doubler votre infrastructure réseau. Si vous avez câblé vos micro-commutateurs en duplex, ce que nous conseillons actuellement (deux fibres, une pour l'émission, l'autre pour la réception), il est possible en changeant juste les deux transceivers SFP d'extrémités de basculer en simplex (émission et réception sur la même fibre optique) une partie ou la totalité de votre architecture. Il est donc possible de doubler votre infrastructure sans intervention lourde et pour un prix modique (2 transceivers SFP LX simplex appariés 1 Gb/s < 2x30 € HT). Ainsi, l'ajout d'un micro-commutateur dans un bureau peut être une opération à coût très faible.

D'autres possibilités existent (micro-commutateur ayant deux ports SFP par exemple) ou n'ont pas encore été découvertes, mais les budgets contraints dans l'ESR (Enseignement Supérieur Recherche) incitent à rester raisonnable dans ses choix. Par exemple, si vous avez 800 micro-commutateurs dans votre projet, un écart de 25 € par élément s'additionne au global en 20 k€ !

**Il faut maintenant éclairer ce réseau.**

Un des problèmes historiques de la FTTO était la concentration optique dans le datacentre impliquant un nombre important de ports SFP voire SFP+ dans les concentrateurs. L'actif était très cher, plus cher qu'un matériel de même performance en cuivre. Cependant, le déploiement massif de la FTTH et FTTB par quasiment tous les opérateurs ainsi que la politique du THD (Très Haut Débit) dans tous les pays ont eu un impact sur le coût du mètre linéaire de la fibre monomode (< 1 € HT/m) et sur le coût des modules SFP désormais fabriqués en très grande série. Ceux-ci sont maintenant accessibles à moins de 20 € HT.

À noter qu'à ce jour, il n'y a pas de haute disponibilité du micro-commutateur. Cependant, la panne d'un équipement ne pénalise qu'un bureau.

## 4  Budget et dimensionnement du projet

*A priori*, de nos jours, le coût d'un câblage en FTTO (ou FTTx [28]) est moins élevé que celui d'un câblage classique en cuivre. À ce jour, la partie passive d'un réseau FTTO est bien moins chère mais l'actif rapporté au port RJ45 est plus cher. Évidement, chaque bâtiment étant unique, seuls des devis chiffrés donnent une réponse claire pour un projet donné. Il est très rare au demeurant sur un projet d'obtenir les factures, de voir les réponses aux appels d'offre. Quelques bâtiments dans l'ESR ont été équipés, ainsi que plusieurs hôpitaux ayant des milliers de micro-commutateurs. Plusieurs entrepreneurs nous ont signalé *oralement* (impossible d'obtenir une trace écrite) que la FTTO avait sauvé les finances du projet ! Il est donc complexe de pouvoir affirmer que la FTTO coûte moins cher pour tout type de bâtiment. Mais il est clair qu'une technologie, qui n'est pas supportée par les principaux leaders du marché de la commutation, ne peut pas prendre autant d'ampleur sans arguments tant techniques que financiers.

Avant de réaliser le cahier des charges, il est nécessaire de définir le plan de déploiement de la FTTO dans votre entité. Il vous faut prendre en compte la volumétrie totale et la durée des travaux (sur plusieurs années, si nécessaire), l'architecture optique de la solution complète, la localisation des éléments actifs et passifs et les différents passages de câble possibles. L'idéal est de fixer tous ces éléments sur un plan de masse qui sera la référence du projet. Selon la nature du projet (neuf ou réhabilitation), le phasage des travaux peut être différent. Lors d'une réhabilitation, il est tout à fait possible de phaser les travaux sur plusieurs années :

1. faire poser la boucle et les boîtiers de raccordement par une entreprise spécialisée. Cette opération est globale sur un secteur ;
2. raccorder chaque bureau par la pose d'une jarretière optique entre un boîtier et le micro-commutateur.

Cette dernière étape peut être réalisée en interne par des personnels BAP G de l'unité et ainsi valoriser leur travail. Contrairement à une installation cuivre, une installation optique FTTO est tout à fait adaptative. Cette architecture permet d'évoluer au fur et à mesure des besoins et des budgets. Le remplacement des liaisons cuivre par des commutateurs optiques dans chaque bureau peut aisément se faire « au fil de l'eau ». Il est aussi possible de prévoir dès le départ des extensions possibles (ou marges de réserve) au niveau d'un groupe de bureaux.

Nous savons que le cheminement du câble optique occupe peu de place et s'adapte au maximum aux chemins de câbles existants du fait de son insensibilité aux perturbations électromagnétiques. Il faut rapidement faire le choix du type de connectique – duplex vs simplex – car cela dimensionne le câble à poser. La position et le nombre de boîtiers de raccordement sont aussi importants. Il faut également faire le choix de la redondance ou non du chemin. En effet, en cas de redondance, deux commutateurs voisins devront être raccordés à deux boîtiers différents. En complément, il est important de garder dans chaque boîtier une réserve en fibres et dans le câble une réserve en tubes. Une rupture accidentelle ou une extension future est toujours possible. Il est admis de conserver 20 à 30 % de réserve. Le choix d'un câblage en duplex des ports SFP permet, en cas de bascule vers du simplex, de doubler son infrastructure. C'est une « assurance-vie » intéressante sur ses premiers projets. Il est en effet plus facile en début de projet de demander du duplex que de demander la pose d'une nouvelle boucle complémentaire quelques années après !

Toujours concernant le câblage, vous devez choisir entre de la fibre monomode OS1 [29] ou OS2. À ce jour, tous les sites sauf un ont choisi l'OS1, dont les performances sont *a priori* suffisantes au vu des distances dans un bâtiment et de sa durée de vie estimée (> 40 ans). De même, vous devez choisir entre une connectique LC ou SC, mais surtout entre UPC (connecteur bleu) et APC [30] (connecteur vert). Avec l'APC, les fibres sont coupées en biais et il y a un peu moins de pertes optiques au niveau des connectiques, mais le prix était, jusqu'à présent, un peu plus élevé. Vous devez alors gérer un stock de jarretières différentes, par exemple des LC /APC – LC/UPC et des LC/UPC – LC/UPC, car les transceivers SFP sont actuellement tous au format LC/UPC. À vous d'estimer le rapport coût performance lors du montage financier de votre projet.

### 4.1 Le cœur de réseau

Le choix du cœur de réseau optique est important. L'idéal est d'en limiter le nombre sur votre site. Selon la criticité, il faut choisir d'en avoir un ou deux. Ce cœur de réseau peut être dissocié d'une salle serveur. Il n'est pas censé héberger de commutateurs PoE, donc son système de refroidissement peut être limité. Il vaut mieux séparer les commutateurs et les bandeaux de brassage dans des baies réseaux dissociées. Le matériel actif dans ce cœur doit absolument accepter les transceivers génériques. Il existe au marché UGAP et Matinfo, ou hors marché, des matériels à 24 ou 48 ports SFP (SFP+), pouvant se stacker, à un prix tout à fait raisonnable.

### 4.2 Les micro-commutateurs

Par ailleurs, il convient de décider si la fonctionnalité PoE doit être présente sur tout ou partie des micro-commutateurs. Sans PoE, un micro-commutateur peut être directement branché sur le 230 V et consomme alors environ deux fois moins que derrière un transformateur 54 V logé en amont dans la goulotte. En effet, le commutateur transmet la puissance PoE mais ne peut pas, vu son faible volume, baisser de lui-même la tension. Le transformateur rajoute environ 50 € HT au prix de base. Globalement, l'idéal est de brancher les micro-commutateurs sur un circuit 230 V protégé (ondulé ou tout au moins un départ électrique dédié). À vous de voir s'il est judicieux de mettre en place ce courant protégé ou s'il est plus simple de connecter les commutateurs sur les prises les plus proches ! À noter que les micro-commutateurs ont été conçus initialement à partir de commutateurs industriels. Ils sont donc durcis avec un MTBF [31] estimé sur certains modèles à plus de 40 ans et optimisent tous leur consommation afin de minimiser la dispersion thermique (fonctionnement en goulotte – espace fortement confiné). Le coût au port RJ45 est plus élevé que sur les autres commutateurs du commerce. À ce jour, on peut estimer qu'un micro-commutateur ne doit pas dépasser 350 € HT sur un devis. Le micro-commutateur étant posé dans une goulotte, vous devez choisir si vous imposez deux ou trois compartiments à celle-ci sur tout le cheminement dans le bureau.

Enfin, avec un projet FTTO, il n'y a plus besoin de locaux techniques dans les étages pour installer des baies de brassage et *in fine* des surfaces (non nobles) sont économisées. En cas de rénovation, ces surfaces peuvent être affectées à de nouvelles fonctions (photocopieuse…).

## 5  Type de cahier des charges

Dans le cadre d'un bâtiment neuf, les lots « courant fort » et « courant faible » sont souvent réalisés par la même entreprise, en général spécialisée dans les courants forts. Les actifs réseaux peuvent être dans un lot distinct.

Il faut être très vigilant sur le cahier des charges écrit par les bureaux d'études. Ils sont souvent influencés par les gros constructeurs qui leur soufflent une solution clef en main, qui ne correspond pas forcément au besoin local. Il faut avoir une démarche pro-active en étant prescripteur des choix techniques et supprimer toute indication technique non nécessaire verrouillant le choix final sur un seul partenaire. Vous devez suivre la rédaction et valider le cahier des charges si vous en avez la possibilité.

Dans le cadre d'un PPP (Partenariat Public-Privé), il est encore plus compliqué de pouvoir intervenir dans le cahier des charges. Il faut donc être extrêmement vigilant sur le cahier des charges initial ! Il est toujours possible de spécifier que le passif soit dans le PPP et que l'actif reste à la charge du bailleur.

Le cas le plus intéressant est celui où votre entité – vous-même – avez la maîtrise d'ouvrage. Vous allez pouvoir rédiger le ou les cahiers des charges. Il est préférable, afin d'optimiser le coût financier et d'avoir la main sur le choix des actifs, de réaliser deux cahiers des charges : un pour les éléments actifs optiques et un autre pour la partie infrastructure.

- Le premier (éléments actifs) doit tout d'abord décrire l'infrastructure existante et les changements demandés. Il doit également inclure la description technique et les caractéristiques applicatives souhaitées des commutateurs et des modules SFP associés. Les caractéristiques du logiciel de management des éléments actifs doivent être décrites. Le cycle de gestion des mises à jour des firmwares doit être précisé. Il faut également faire clarifier l'étendue de la garantie et du support technique. Enfin, on peut préciser si une formation doit être incluse.
- Le second (partie passive) concerne tous les travaux d'infrastructure : la dépose éventuelle des anciens matériels, la topologie de la future installation, les cheminements et raccordements, la baie réseau, les raccordements dans les bureaux, l'étiquetage, l'organisation des travaux, les garanties et la conformité de la nouvelle installation.

Dans tous les cas, il est souhaitable de demander un PoC (Proof Of Concept) sur site par les constructeurs intéressés, c'est-à-dire de faire réaliser une maquette expérimentale de faisabilité. Dans celle-ci, il faudra notamment être attentif aux protocoles d'authentification des accès réseaux, mais aussi aux protocoles de détection de boucle. En effet ces derniers s'avèrent être le problème le plus difficile à résoudre pour certains constructeurs. D'autres cas pratiques, liés à votre architecture ou à vos flux, peuvent être envisagés. On peut citer par exemple de l'injection de paquets afin de saturer une interface *uplink* tout en continuant à faire fonctionner correctement la ToIP.

Rapprochez-vous du groupe FTTO qui dispose déjà de sites en production et qui vous aidera et vous guidera. Il est préférable d'avoir un ou plusieurs avis extérieurs au projet.

## 6 Type et configuration des (micro) commutateurs

Il existe plusieurs solutions pour administrer les micro-commutateurs. Ce choix va dépendre du nombre d'équipements à configurer. Il se dégage principalement 6 solutions complémentaires :

1. l'interface web HTTPS ;
2. la connexion SSH avec des scripts ;
3. la supervision en SNMP ;
4. le chargement, la sauvegarde de configuration ou la mise à jour en TFTP et FTP ;
5. le mode *standalone* avec l'application du constructeur ;
6. le mode client-serveur avec l'application du constructeur + des scripts.

Les micro-commutateurs sont des matériels actifs d'extrémité. Leurs OS ont été adaptés pour une gestion centralisée. Ils permettent une gestion simplifiée mais fine des paramètres de configuration (LLDP, VLAN, ToIP, Radius, BPDU Guard, STP, Loop Protection, NTP, DHCP Snooping…). L'administration d'un seul ou d'un parc entier de commutateurs est en pratique aussi simple. Pour la configuration initiale, l'idéal est de créer un master et de le déployer.

Pour une gestion dynamique du réseau et des machines, l'authentification peut être réalisée via des serveurs Radius. L'avantage de cette solution est de banaliser l'accès aux prises réseaux dans les bureaux. Les fonctionnalités de *Loop Protection* et de *Spannning Tree* sont également paramétrées et sont essentielles en cas de boucle (redondance de chemin notamment). Pour superviser et monitorer les micro-commutateurs, on peut utiliser une sonde ou utiliser l'application du constructeur.

Comme tout matériel informatique, il est fortement recommandé de faire une lecture attentive des notes de mise à jour (correction de bogues et ajout de fonctionnalités). Ces mises à jour doivent être testées sur quelques équipements puis appliquées en parallèle par groupes d'équipements. Selon les marques, il faut compter entre 5 et 12 min par micro-commutateur.

La fonctionnalité PoE est utilisée pour les bornes Wi-Fi, la téléphonie sur IP et les tablettes. Pour des questions d'économie d'énergie, il est possible d'éteindre ces équipements durant la nuit. Les micro-commutateurs respectent également la mise en veille.

Globalement les solutions sont conviviales, fiables, robustes et simples à administrer avec l'application du constructeur. Les MTBF (entre 100 000 h [32] et 750 000 h [33]) des micro-commutateurs sont élevés à très élevés. Les taux de panne sont donc théoriquement assez bas. En ce qui concerne la maintenance, les micro-codes sont, pour l'instant, fournis à vie. Le logiciel de gestion propriétaire est, lui, gratuit ou payant annuellement selon le constructeur. Au vu du nombre de micro-commutateurs dans les bâtiments, un contrat de maintenance classique (J+1 par exemple) a peu de sens. On préconise donc d'avoir quelques commutateurs en stock, pour parer à une panne le temps d'envoyer le matériel en réparation. Les retours de panne des matériels des deux leaders de la FTTO montrent que ceux-ci sont réellement réparés en Europe et ne font pas l'objet d'un échange standard. Ceci est très positif pour l'économie circulaire.

À des fins de développement durable (critère à 5 % dans de nombreux marchés publics [34]), il convient de spécifier dans un cahier des charges le MTBF minimal souhaité (par exemple 20 ans). Il est possible de demander l'engagement du constructeur sur la durée des mises à jour et la réparabilité des produits après l'arrêt de la production d'un modèle. Il est aussi intéressant de faire remplir un tableau avec la consommation électrique avant et après le bloc d'alimentation pour les trois configurations suivantes : en veille, avec un équipement branché en RJ45, à pleine charge.

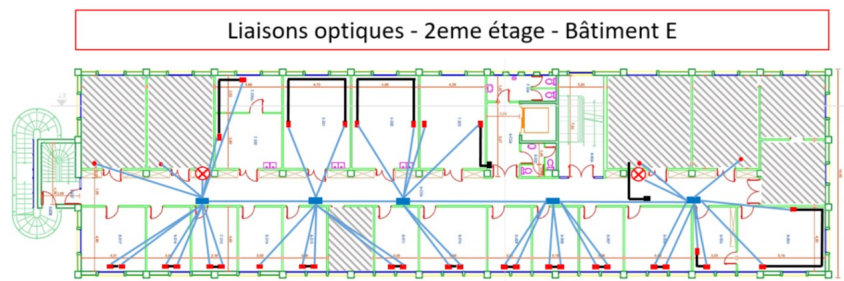

*Figure 14: Plan de câblage mixe FTTO / classique avec raccordement des anciens réseaux sur la boucle optique*

À noter enfin que la configuration d'un réseau FTTO est 100 % compatible avec un réseau classique. Il est ainsi possible dans une grande pièce ou proche d'une expérimentation d'avoir un commutateur administrable 24 ports (figure 14) ! Il est aussi envisageable de mettre des commutateurs sur table, permettant juste une conversion cuivre/optique et centraliser l'intelligence sur le cœur de réseau. On peut clairement se poser cette question quand on doit, par exemple, distribuer une salle de cours informatique dans une école.

## 7  Conclusion

Nous venons de faire un tour d'horizon très rapide de la technologie FTTO et de quelques retours d'expériences au sein de l'ESR. Les avantages de l'implantation de la fibre jusque dans les bureaux sont multiples :

- un débit en monomode évolutif. Rien ne s'oppose plus aux 10 Gb/s, 25 Gb/s, 100 Gb/s, 1 Tb/s… ;
- une absence de la sensibilité aux interférences électromagnétiques. Comme corollaire, votre architecture bénéficie de l'isolation galvanique entre le cœur de réseau et les bureaux ;
- la disparition des répartiteurs d'étage ;
- des chemins de câble considérablement réduits. Ils sont partagés avec les autres réseaux (courant fort, SSI [35], GTB, contrôle des accès…) et impliquent un gain de place dans les faux plafonds ;
- une pose très rapide des boucles optiques. Quelques opérations nécessitent un ouvrier spécialisé (dérivation, fusion, recette optique). Le reste est facile à réaliser sans compétence particulière ;
- une évolutivité de l'installation sans effectuer de gros travaux. Il est ainsi possible de migrer d'un réseau cuivre vers un réseau optique, bureau par bureau, sans interrompre la production dans les pièces voisines ;
- un coût en baisse chaque jour et de plus en plus attractif.

La FTTO s'inscrit aussi dans une démarche de Green IT et d'économie d'énergie [36]. Elle est très légère (gain en poids) et ne consomme quasiment plus de cuivre, ressource qui se raréfie [37]. Les très faibles longueurs en cuivre (redondance de chemin, jarretière terminale, *etc*.) induisent moins de pertes notamment en PoE. Les locaux techniques souvent climatisés n'ont plus lieu d'être (gain de m², moins de béton). Enfin, la durée de vie estimée des fibres optiques est actuellement de 40 ans.

**Alors, êtes-vous prêt à franchir le pas ?**

Cette solution appelée FTTO a été mise en place, en 2013 au LEGI de Grenoble. D'autres entités ont suivi, chacune avec ses spécificités. Ce texte a voulu présenter et dégrossir les points importants afin que les projets FTTO à venir aient le plus de cartes en main pour être menés à bien.

L'expression « **tuer le cuivre** » est un peu violente, mais si cela a éveillé en vous de l'intérêt et de la motivation, notre pari est gagné.

## Annexe : Repérage et numérotation des fibres, boîtiers de dérivation et micro commutateurs optiques

Attention, cette partie est en annexe, mais elle est cependant très importante, fondamentale ! Les fibres sont posées pour 40 ans, donc votre repérage doit pouvoir être transmis de personne en personne, être compris par les BAP G (Infrastructure / Bâtiment) et les BAP E (Informaticien ASR) de votre génération et de la suivante.

Il ne faut pas hésiter à mettre des étiquettes sur le panneau de brassage central, sur chaque boîtier de raccordement et sur une partie fixe de la goulotte proche du micro-commutateur. Il faut cependant éviter l'unique étiquette collée sur le commutateur qui part avec celui-ci en maintenance et ne revient pas… Des bonnes étiquettes doivent durer et coller plus que 5 ans !

Il existe un **code couleur** au niveau des câbles et des brins optiques qui facilite le repérage. Chaque fibre et chaque tube a une couleur. Les constructeurs utilisent tous la même palette. Un code couleur permet donc de numéroter les tubes et les fibres dans les tubes. Il en existe plusieurs, mais le code FOTAG **[38]** est le plus utilisé dans les réalisations FTTO actuelles.

| Numéro | Alphabétique | FOTAG (IEEE 802.8) | France Télécom |
|---|---|---|---|
| 1 | Blanc | Bl/Blu – Bleu | RO/Rouge |
| 2 | Bleu | Or/Orange | BE/Bleu |
| 3 | Gris | Gr/Green – Vert | VE/Vert |
| 4 | Jaune | Br/Brown – Marron | JA/Jaune |
| 5 | Marron | Sl/Grey(Slate) – Gris | VI/Violet |
| 6 | Noir | Wh/White – Blanc | BC/Incolore (Blanc) |
| 7 | Orange | Rd/Red – Rouge | OR/Orange |
| 8 | Rose | Bk/Black – Noir | GR/Gris |
| 9 | Rouge | Yl/Yellow – Jaune | MA/Marron |
| 10 | Turquoise | Vi/Purple – Violet | NO/Noir |
| 11 | Vert | Pk/Pink – Rose | TU/Turquoise |
| 12 | Violet | Tu/Turquoise | RS/Rose |

Il est important de se renseigner sur le site afin de savoir quel est le code couleur choisi au niveau du campus. Il faut imposer le code couleur à utiliser dans votre cahier des charges afin que toute votre installation suive la même nomenclature.

*JRES 2019 – Dijon* 16/19

**Exemple de repérage complet :**

Le repère du type de commutateur comprend : le sens du câble / N° câble / N° tube / N° fibre

- Sens du câble : **A** ou R (Allé ou Retour)
- Numéro du câble : A, B, **9** (lettre ou chiffre)
- Numéro du tube : **0**, 1, 2, 3, 4, 5, 6, 7, 8, 9, A, B
- Numéro de la fibre : **0**, 1, 2, 3, 4, 5, 6, 7, 8, 9, A, B
- Code couleur : FOTAG
- Repère boite de raccordement FO : A9-0
- Repère type de commutateur : A-9-0-0
- Localisation du bureau : 4-A-001
- Origine depuis le datacentre :
    - Baie : MA-2
    - Stack : S3-1
    - Port : 30
- Extrémité depuis le datacentre :
    - Baie : BC-2
    - Tiroir : T13
    - Port : A9-0-0

**Exemple, cas du LEGI :**

- Code couleur : FOTAG
- Numéro de la boucle : 1, 2 ou 3 (3 câbles de 12 tubes de 6 fibres)
- Sens du câble : A ou B
- Nom des boîtiers : BR102 (numéro boucle, numéro boîtier sur la boucle à deux chiffres)
- Tube sur boîtier : A-Bleu, B-Rouge (sens et nom de la couleur des tubes dérivés)
- Port sur boîtier : 1 à 12 par paire, d pour duplex, a ou b selon le port simplex
- Sur panneau de brassage : BR102.1 (nom du boîtier, premier port de la série)
- Sur micro-commutateur : BR102.1d (duplex), BR102.2a, BR102.2b (simplex)
- Nom du commutateur (DNS) : sw-legi-k213-b1 (k213 : nom du bureau, b : bureau, c : couloir – on met alors le nom du bureau le plus proche)

Dans le serveur de nom, un commentaire permet de faire le lien entre le nom DNS du commutateur faisant apparaître le bureau et le nom technique permettant de retrouver la topologie. Tout est en parallèle documenté dans un wiki. La redondance d'information permet parfois de corriger des erreurs.

# Bibliographie / Lexique

## Licence

Licence d'utilisation : ***Licence Art Libre version 1.3 ou supérieur*** (http://artlibre.org/).

Cette licence est équivalente dans l'esprit à la licence GPL version 2 ou supérieur conçu pour le code source. C'est une licence à gauche d'auteur.

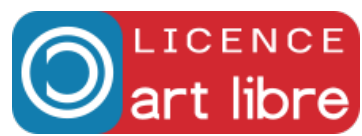